\documentstyle[11pt,aaspp4,psfig]{article}

\begin{document}
 \title{  Gravitational Lensing by Cold Dark Matter Catastrophes}
\author{Craig J. Hogan}
\affil{Astronomy and Physics Departments, Box 351580, University of Washington,
    Seattle, WA 98195}
\date{\today}
\def\msol{{\,\rm M_{\odot}}}
\begin{abstract}
Intrinsically cold particle dark matter inevitably creates
halos with sharp discontinuities in projected surface density
caused by the projection of fold catastrophes onto the sky.
In principle, these imperfections can be detected and measured with 
gravitational lensing through discontinuities in image magnification
and image structure. Lens solutions are discussed for  the most common
universal classes of discontinuities. Edges caused by  cold particles 
such as  condensed axions and thermal WIMPs are very sharp, respectively
about $10^{-12}$ and $10^{-7}$ of the halo scale. Their structure can be
resolved by stellar and quasi-stellar sources which show sudden changes
in   brightness or even sudden disappearances (sometimes within hours) as
edges are crossed. Images of extended objects such as edge-on galaxies 
or jets can show sudden bends at an edge, or  stretched, 
parity-inverted reflection symmetry about a sharp line. 
Observational strategies and prospects are briefly discussed.
 
\end{abstract}
\section{Singularities in Projected Density}
Particle dark matter naturally forms thin structures in 
phase space (\cite{ring}). In some cases, such as axions   forming by
Bose condensation, the peculiar velocities are exceptionally
small, but even for thermally produced particles,  adiabatic cooling
creates a cool Hubble flow which resembles a 3D
sheet ($\vec v= H\vec r$) in six-dimensional phase space. The fine-grained phase
structure remains thin even after nonlinear collapse destroys
the ordered Hubble flow.  For systems which are not too
dynamically old or too violently relaxed, the fine structure
retains information about the nature of the dark matter and its dynamical history.

The fine-grained structure remains most conspicuous at the locations
of caustics in the flow of matter.
 The most famous examples in cosmology
are the Zel'dovich ``pancakes'' (\cite{zeldovich}).
Similar caustics occur
 in general
nonlinear collapses in smooth potentials: for example, shells in spherical
similarity solutions (\cite{fillmore,bertschinger}),
 or ringlike tubes in collapses having
less symmetry (\cite{ring}).
Singularities are not an artifact 
of the symmetry: fold catastrophes (2D sheets) joining in cusp catastrophes
(1D lines), which themselves join in various pointlike
catastrophes,  are
a generic feature of any smooth 3D phase sheet mapping onto
3D space (\cite{arnold,tremaine}). Physically these structures are caused
by turning points or focus points of orbits.  They create formal singularities
in density (sheets, lines and points) with universal
density profiles.  These   features may be observable through their
dynamical effects (\cite{rings}), though with insufficient resolution to reveal
the fine structure of the dark matter phase sheet.

Here I discuss the projection of the phase sheet onto
the sky and how   discontinuities in surface density
can manifest themselves in gravitational lensing. Lensing offers
the advantage of very small background sources which can serve
as microscopic probes of small scale phase structures. Observations
of such events can yield physical information about the nature
of the dark matter and its formation mechanism-- for example,
whether axions formed by condensation or   string
radiation.

The most common generic true spatial catastrophes of cold dark  matter---
 a 3D phase sheet moving in a 3D 
subspace of 6D phase space--- are ``folds'', 2D structures of codimension
and corank 1 (\cite{tremaine}). 
 They terminate on linelike cusps which in turn 
terminate on pointlike swallowtail, elliptic umbilic and
hyperbolic umbilic catastrophes.
 Folds form sheets with   a formally divergent  density,
with a universal density
profile: on one side   $\rho\propto  x^{-1/2}$,   abruptly falling to $\rho=0$
on the other side.\footnote{Note that these structures actually 
exist in the fine-grained halo density distribution. In principle, a
 detailed
time history of the WIMP flux on Earth would show spikes of divergent
flux every few million years. The integrated flux in these spikes
however is small  so in a real fossil record (such as tracks
in mica) they would not stand out clearly against the background.}
  However, when projected onto the sky
the catastrophic divergence does not survive. Integrating
the fold profile in projection produces instead a 
 a
simple jump in column density, with a sharp edge, along the line  
where a fold sheet is convex and  tangent to the line of sight.

Consider a (generic) 2D fold of a 3D phase sheet in 3D space. Let the $z$ axis lie
along the line of sight and tangent to the fold, the $y$ axis in the plane of the
sky and tangent to the projection of the fold at $x=0$. The sheet lies at
$x_s(z)=z^2$, so the 3D density
to the right of the sheet is $\rho= (x-z^2)^{-1/2}$.
The projected density is
\begin{equation}
\rho(x)=\int_{z=-\sqrt{x}}^{z=+\sqrt{x}}\rho(z')dz'=
\left\{ \begin{array}{ll}
\pi &\mbox{for  $x>0$}\\
 0&\mbox{for $x<0$,} 
\end{array}
\right.
\end{equation}
a discontinuous jump where the projected fold is tangent to the line of sight.
 The projected edges  form lines on the sky with coherence
scale $\ell$, along which the amplitude of the discontinuity slowly varies. The
amplitude of the jump (which depends on the curvature of the fold
in the invisible dimensions), and its coherence scale on the sky,
 depend  on the  
  dynamical substructure and history
of the halo.
Roughly speaking, a ``smooth'' halo
which is $N$ orbital times old
has about $N$ wraps and folds of its phase sheet, and
about $N$ projected edges. 
 A typical discontinuity
has an amplitude $\delta\Sigma \approx N^{-1} \bar\Sigma$
where the  mean halo surface density is $\bar\Sigma$.

In reality the catastrophes are  concentrated in the central,
dynamically old parts of a halo and display a chaotic structure
reflecting the dynanical history of the system; only the younger
folds, representing particles on their first or second orbits, show
coherence
on the scale of the halo itself.  Although the 
statistical properties of the edges in realistic
models have not been studied in detail, they are apparent in 2D projections
of dynamically young systems in high-resolution N-body simulations
(J. Stadel, private communication).
Edges disappear if the matter is too hot or if the experiment
is too coarse-grained compared to the scale of the dynamical wraps.

  At the finest resolution, the sharpness of the edge
 discontinuities  is in principle limited only by the intrinsic width of the 
dark matter phase sheet.
This scale is typically  
 much smaller than the angular scale $\ell$ of the dynamical folding,
in which case the lensing effects can be treated as a one-dimensional problem.

\section{ Lensing by Projected Folds}

\subsection{Lenses with 1D Translational Symmetry}
 We
 follow the  notation   of 
Blandford and Narayan (1986), 
describing relativistic bending in the thin-screen approximation
by a two-dimensional potential $\psi$
on the plane of the sky $(x,y)$.
 We study line discontinuities on the $y$ axis and
because of the translational symmetry
 the 
 two-dimensional Poisson equation   reduces to
one dimension,
\begin{equation}
{\partial^2\psi\over\partial x_I^2}=2{\Sigma(x_I)\over\Sigma_c}.
\end{equation}
The critical surface density 
 $\Sigma_c=c^2(4\pi G D)^{-1}= 0.35 {\rm g\ cm^{-2}}(D/1{\rm Gpc})^{-1}$ is 
determined by the  combination $D=D_d D_{ds}/D_s$
of the  angular diameter distances to the lens $D_d$ and the source
$D_s$, and between them $D_{ds}$;
at this density 
a uniform sheet of matter would just focus radiation
from the source at the observer.
  The positions
of the images $x_I$ are determined by the lens equation
\begin{equation}
x_I=x_s+{\partial\psi(x_I)\over\partial x_I}
\end{equation}
where $x_s$ is the direction of the source.  
The image distortions and magnifications are
determined by the extrinsic curvature tensor
\begin{equation}
K_{ij}={\partial\theta_{Si}\over\partial\theta_{Ij}}=
\left[
\begin{array}{cc}
\kappa+\mu & 0\\
0 & \kappa-\mu
\end{array}
\right]
\end{equation}
where the principal axes for 
lensing are the same as those set by the discontinuity,
 $\theta_1=x$ and $\theta_2=y$.
Here
$\kappa=1-(\Sigma/\Sigma_c)$
 denotes the expansion,
and $\mu$ denotes the shear associated with 
image distortion. By symmetry in this situation,
the $yy$ term of $K$ is unity, determining
a  particular relation between expansion and shear:
$\kappa-\mu=1$,
 and hence 
$\mu=-(\Sigma/\Sigma_c)$.
The magnification of an image is given by
$M=|\kappa^2-\mu^2|^{-1}$ and hence
\begin{equation}
M=\left|1-2{\Sigma\over\Sigma_c}\right|^{-1};
\end{equation}
images are stretched in the $x$ direction by this factor.

\subsection{Folded 2D phase sheet}
Some dissipative  
structures (such as cold disks)   resemble
a 2D  phase sheet, whose projection onto the sky can 
create  high density features resembling   a  
fold catastrophe in projection
(for example, stellar
 shells around elliptical galaxies;  \cite{quinn,hernquist}). In reality these
 sheets are not as thin as intrinsically 3D dark matter
 phase structures although they may produce  recognizable 
 pseudo-catastrophe events.

  In the plane of the sky consider a  fold  
on the $y$ axis at $x=0$. The universal surface density profile   is
\begin{equation}
{\Sigma(x)\over\Sigma_c}=\left\{ \begin{array}{ll}
 x^{-1/2} &\mbox{for  $x>0$}\\
 0&\mbox{for $x<0$} 
\end{array}
\right.
\end{equation}
We have used the scaling freedom to set $\Sigma=\Sigma_c$
at $x=1$.
The  solution of the lens equation yields the mapping between source and image 
positions,
\begin{equation}
x_I 
=\left\{ \begin{array}{ll}
 (2\pm\sqrt{4+x_S})^2&\mbox{for $x_I>0$}\\
  x_S  &\mbox{for $x_I<0$} 
\end{array}
\right.
\end{equation}
Three images in a line are 
produced for $x_S$ in the vicinity of the fold,
$-4<x_S<0$.
 The magnifications  of the images
at $x_I>0$ are
\begin{equation}
M_{\pm}=|1-2x_I^{-1/2}|^{-1}
=\left|1\pm \left(1+{x_S\over 4}\right)^{-1/2}\right|.
\end{equation}
These formulae describe a universal family of solutions
characterized by one parameter. However, the approximation 
breaks down  on a 
scale determined by the temperature of the 2D phase sheet; while
it might serve as a good description of some baryonic  lenses, we
cannot use this as
a probe of dark matter.

\subsection{Sharp Edge from Folded 3D Phase Sheet}

The more interesting case is
 the projection of generic catastrophes caused by folds
of the very cold 3D CDM phase sheet, which create
discontinuities with extremely sharp edges, and
a projected profile 
\begin{equation}
{\Sigma(x)\over\Sigma_c}=\left\{ \begin{array}{ll}
\sigma_0 &\mbox{for  $x>0$}\\
 0&\mbox{for $x<0$.} 
\end{array}
\right.
\end{equation}
The lensing solution is simply
\begin{equation}
x_I 
=\left\{ \begin{array}{ll}
x_S/(1-2\sigma_0)&\mbox{for $x_I>0$}\\
  x_S  &\mbox{for $x_I<0$} 
\end{array}
\right.
\end{equation}
There are no multiple images for $2\sigma_0<1$.
The mapping from $S$ to $I$   receives a stretch in the $x$ direction
to the right of the edge.  There is an abrupt change in 
magnification  from one side of the edge to the other,
from $M=1$ to $M=|1-2\sigma_0|^{-1}$,  which would 
appear  as a sudden change in the brightness  of a small moving
source.
The discontinuity can also  be observed in image structure
 if there is an extended background
source (such as a compact jet or edge-on disk) which lies in a straight
 line:
at the location of the edge the image displays a sudden
bend.  

If $2\sigma_0>1$, two images form if $x_S<0$ (at $x_I>0$
and $x_I<0$) and no images form
if $x_S>0$.  A   source moving towards
the edge from $x<0$  appears as two images approaching each other
with a uniform magnification (one image with $M=1$, the other
with $M=|1-2\sigma_0|^{-1}$, both moving with angular velocity
$\propto M$)  until
it reaches the edge, when the images
meet and both suddenly disappear. The image of an extended object
at $x_S<0$
appears reflected across the line, stretched  and parity-reversed;
that is, the surface brightness at
$(x,y)$ is duplicated at $(-Mx, y)$. 

These very simple solutions only apply over angles much smaller than $\ell$; 
particularly in the supercritical case there are nontrivial macroimages with
a larger-scale structure not described here. However, these solutions 
accurately  illustrate the unique
effects near the edges.

\section{Observing the Innermost Phase Structures}

Physically, the edge discontinuity is the most interesting case since
the width of the edge reveals the initial dark matter
velocity dispersion. An extreme example is  condensed axions, with
 dispersion at the redshift of  collapse $z_{coll}$   (\cite{ring})
\begin{equation}
\delta v_a/c\approx  3\times 10^{-17} (m_a/10^{-5}eV)^{-1} (1+z_{coll}).
\end{equation}
A more typical case is thermally produced  WIMPs which   have
\begin{equation}
\delta v_W/c\approx    10^{-11} (m_W/GeV)^{1/2} (1+z_{coll}).
\end{equation}
This dispersion creates a range of energies at a given initial position,
translating after one orbit to
a dispersion of
turnaround radii for the first fold and smearing  the projected edge
to a  fraction of the halo  size  
$\approx \delta v/v_{vir}$ where $v_{vir}$
 ($\approx 10^{-3}c$)  is the halo
dispersion. This ratio is very small---  adopting
 $z_{coll}\approx$ 3 to 10 typical of CDM models, 
it is about $10^{-12}$ for axions and $10^{-7}$ for WIMPS.
If the edge results from a fold of angular size $\ell$,
its  angular width is about 
 $\delta x \approx \ell \delta v/v_{vir}$. 
 The edges are therefore very 
sharp,  ranging  from a few  milliarcsec for WIMP edges in our local
halo ($\ell\approx 0.1$, $\delta x\approx 10^{-8}$))
 to a few picoarcsec for axion
 edges in clusters or halos  at high redshift ($\ell \approx 10^{-4}$,
$\delta x\approx 10^{-16}$.

Resolving the structure in  these extraordinarily sharp edges---
the interesting goal if we want to use them to probe dark matter---
 requires small (and hence faint) background sources.
For  halo-scale folds ($\approx 10^{23}$ cm),
normal solar-type  stars ($\approx 10^{11}$ cm) are about the
right size to resolve the axion edge.
Bright sources such as   quasars and supernovae
 ($\approx 10^{15}$ cm) are too large to resolve the axion
edge but have the advantage that they are  visible at high
redshift, and  can still
  resolve edges produced by thermal WIMPs. 
In principle at least,  realistic background sources can 
probe the initial dark matter distribution function.

How practical is it to use these occurrences as a probe?
One type of experiment would be MACHO-style (\cite{macho,pac96}):
 monitor a large 
number of background sources (such as a galaxy of
stars) and watch for   sources crossing edges in a foreground
halo. Clearly, the light curves caused by 
edge transits are very different from the ``standard'' MACHO events
of point-mass microlensing (\cite{paczynski}).
The number
of edges across the face of a halo is about the number of orbits
since it formed, say $N\approx 10$ to 100. Each background star traverses
the angular scale of the halo in a dynamical time $t_{halo}\approx 10^8$ yr
so monitoring $10^8$ stars for a year will show 10 to 100 of them
crossing edges.
The timescale for the brightness to change
(assuming the source resolves the edge)  is the  transit time
 of the edge 
$\approx t_{halo}(\delta v/v_{vir})$, about
an hour for the axion edge and years for thermal
WIMPs.   

 Unfortunately  the amplitude  $\sigma_0\approx \bar\Sigma/N$
 of the  
jumps  is less than
$10^{-6}$ for Local Group experiments, which is not
observable given limitations of   photometric accuracy 
 and the stability of the background sources.     
 To find large-amplitude or ``appearing/disappearing'' events,
sources must be monitored at high redshift where 
$\sigma_0$ in typical halos
can approach unity.

 This requires surveys which
are simultaneously larger, deeper, and better sampled in time
than is possible with current telescopes. If the goal is to
resolve the axion edge, one must detect a solar-size star
in an hour at the Hubble length, which takes a 1000m telescope
(fortunately, this  is also large enough to resolve a galaxy this far away
into stars). With a 100m telescope--- which is at least  being 
contemplated--- the experiment could be done with giant stars
as targets, the events lasting of the order of days.

In the short term it
 is more realistic to study these effects via imaging.
Discontinuities can be found by sharp bends  in  images of  extended linear
objects, such as jets or edge-on galaxies.  Imaging of
jets with  mmVLBI (with resolution potentially as high as
$\delta x\approx 10^{-10}$) or VLBA  
($\delta x\approx 10^{-8}$)  can
verify the presence of sharp edges and test for the cold
character of the primordial phase sheet.
Although the accessible angular scales are not nearly small enough
to probe axions, they can reach the WIMP scale and test for
warmer forms of dark matter.

Most interesting are
high amplitude supercritical edges ($2\sigma_0>1$) 
in dense halos and clusters at high redshift with
$\bar\Sigma>\Sigma_c$. They create double images 
of extended objects, reflected about a sharp line and stretched 
perpendicular to the line. A jet would not only appear
to bend sharply, it would appear the same (but mirror-reversed) on the two sides
of the bend except for a constant stretch factor
in $x$ (and no stretch in $y$). The sharp character
of the edge is apparent if the  resolution is better than the coherence
scale
$\ell$ of the folds, which  may   be almost as large as the macrolens
itself in
dynamically young systems. A fresh fold in a young cluster,
  many arcseconds across, could reflect  a whole galaxy image.
   Edge lensing might 
even appear in high-resolution images
of  galaxies and quasars in known macrolenses (e.g.
\cite{castle}). Note that
lensing conserves surface brightness so there must be  image structure on
scales below $\ell$ for the edge to be visible.

 Even though it may be possible
to resolve 
 the predicted scale of the thermal
WIMP  candidates,  we will have to get lucky to 
find a source this small lying right behind an edge.
One can imagine searching for jets behind the small fraction of the
sky (about $10^{-3}$) with supercritical surface density in a
foreground
halo. Suppose we are looking at jets with a size $x_j$; a fraction
$\approx x_j/\ell$ will happen to transect a young edge in the halo.
This means that about a fraction  $10^{-3}x_j/\ell$ of all jets
will be suitably placed; if we imagine that there are $10^8$ 
jets on the sky (judging from the total number of galaxies and
their duty cycle as AGN), we ought to be able to find 
$10^5x_j/\ell$ of them (times whatever aspect ratio
they have)  which probe edges on the scale
$\delta x\approx x_j$--- or about $10^{-2}$ for WIMPs.
 So, it is realistic to expect to probe well below
the halo dispersion for the coldness of the dark matter,
 but probably not reasonable to test models
of WIMPS. 

The ultimate goal of using this effect to probe the 
primordial phase structure of particle CDM is therefore still
some way off. However,
relatively coarse resolution  still offers a test of
scenarios designed to 
smooth halo substructure or soften halo cores
  (e.g., \cite{ghigna98,moore}): for example, reducing small-scale
fluctuation power   would increase  the fold coherence scale,
and raising $\delta v$ would  increase  the thermal
 smearing scale, ultimately 
eliminating the edges altogether. Observation of  a sharp dark matter
edge on sub-halo scales could rule out hot dark matter candidates.
Although the theoretical predictions need to be sharpened, it
seems likely that
some of these  tests can be performed using current technology,
even using background galaxies as sources, with a sufficiently
large survey.

\acknowledgments
I am grateful to S. Phinney,  G. Lewis, C. Stubbs, D. Brownlee,
and S. D. M. White for useful comments and discussions.
This work was supported by NASA  at the
University of Washington, and by a Humbolt Research Award
at the Max Planck Instibute f\"ur Astrophysik.

\begin{figure}[t]
\vspace{0.5in}
\centerline{\psfig{file=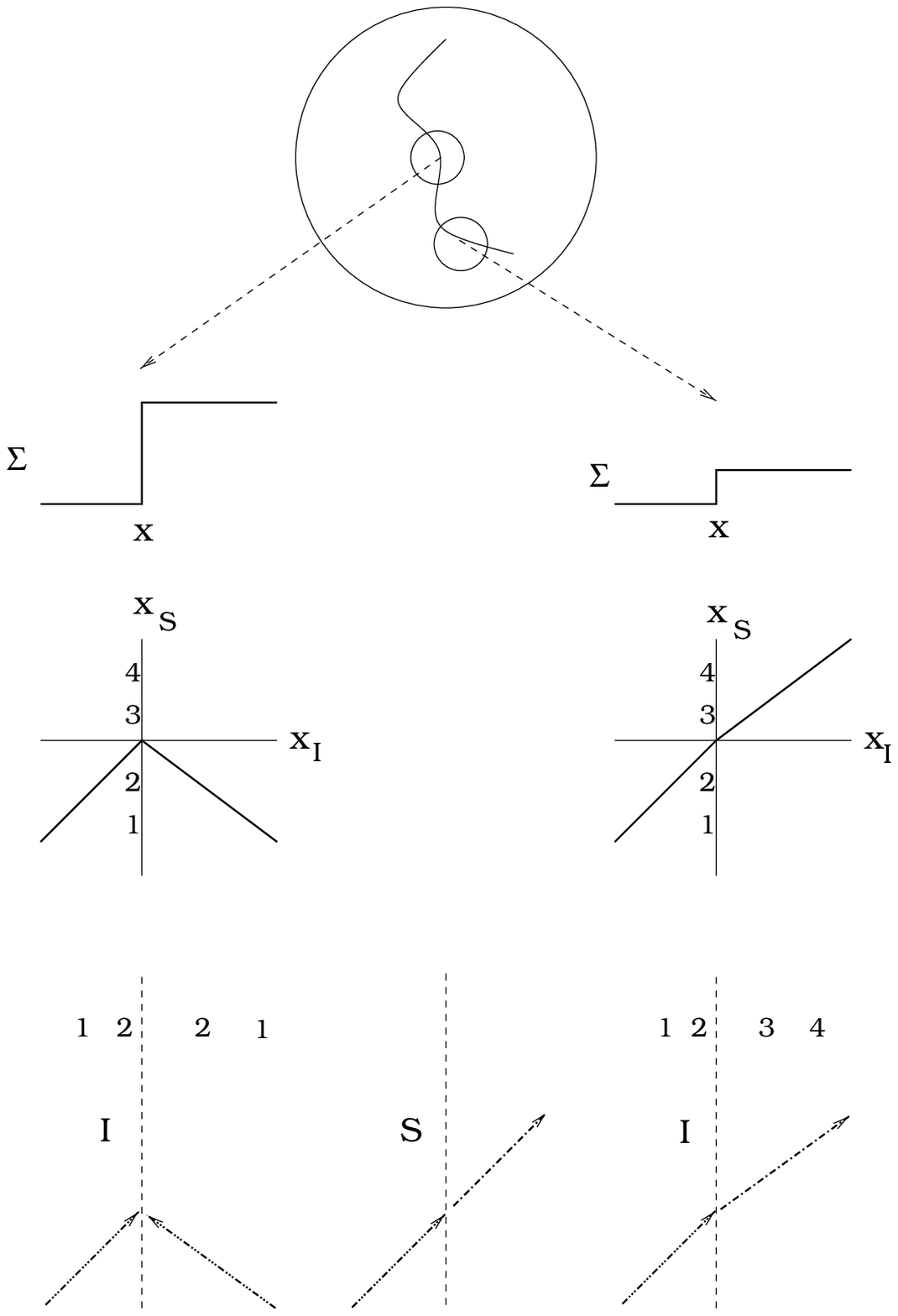}}
\vspace{0.5in}
\caption{Illustration of lensing by a fold catastrophe.
At top, a halo has an edge (one of many)  with angular size $\ell$.
A surface density cut across the edge
is shown at two different locations, with different amplitudes.
The graphical solution of the lens equation is shown for 
supercritical (left) and subcritical (right) cases, showing
the mapping of the source position to image position.
The appearance on the sky is shown at the bottom;a moving
source is shown at four times, showing its disappearance 
in the supercritical case; an arrow shows a reflected/stretched and
bent image respectively  in the two cases.
}
\end{figure}

\end{document}